\documentclass[oneside,english]{amsart}
\usepackage{amssymb}

\makeatletter

 \theoremstyle{plain}
 \theoremstyle{plain}    
 \newtheorem*{thm*}{Theorem} 
 \theoremstyle{plain}    
 \newtheorem*{lem*}{Lemma} 

\newcommand{\be}{\begin{equation}}
\newcommand{\nd}{\noindent}
\newcommand{\ee}{\end{equation}}
\newcommand{\ben}{\begin{eqnarray}}
\newcommand{\een}{\end{eqnarray}}

\usepackage{babel}
\makeatother
\begin{document}

\title{Stochastic invertible mappings between power law and Gaussian probability distributions }

\author{C. Vignat$^1$ and A. Plastino$^2$\\\\$^1$ L.I.S. Grenoble, France and E.E.C.S., University of Michigan, USA\\ e-mail: vignat@univ-mlv.fr  \\\\ $^2$ La Plata National University $\&$ National
Research Council (CONICET) \\ C. C. 727 - 1900 La Plata -
Argentina \\ e-mail: plastino@fisica.unlp.edu.ar  }

\begin{abstract}

We construct ``stochastic mappings" between power law probability distributions (PD's) and Gaussian ones.  To a given vector $N$, Gaussian distributed (respectively $Z$, exponentially distributed),  one can  
 associate  a vector $X$, ``power law distributed", by multiplying $X$ by a random scalar variable $a$, 
 $N= a X$. This   
 mapping is "invertible": one can go 
 via  multiplication by another random variable $b$ from $X$ to $N$ (resp. from $X$ to $Z$), i.e., $X=b N$  (resp. $X=b Z$). Note that  all the above equalities 

 mean "is distributed as".  
  As an application of this stochastic mapping we revisit the so-called ``zero-th law of thermodynamics problem" that bedevils 
  the practitioners of nonextensive thermostatistics.
 
 
 \end{abstract}
\maketitle
\section{Introduction}

\nd Statistical Mechanics' most notorious and renowned probability distribution (PD) is that deduced by Gibbs for the canonical ensemble \cite{reif, pathria}, usually referred to as the Boltzmann-Gibbs (BG) equilibrium distribution

\be \label{gibbs}  p_G(i) =\frac{\exp{(-\beta E_i)}}{Z_{BG}}, \ee with $E_i$ the  energy of the microstate labeled by $i$, $\beta=1/k_B T$ the inverse temperature ($T$), $k_B$ Boltzmann's constant, and $Z_{BG}$ the partition function. The exponential term
$F_{BG}=\exp{(-\beta E)}$ is, of course, called the BG factor. Recently Beck and Cohen 
\cite{super} have advanced a generalization (called ``superstatistics")  of this BG factor, assuming that the inverse temperature $\beta$ is a 
stochastic variable. They effect a multiplicative convolution 
to obtain a generalized statistical factor 
$F_{GS}$ in the fashion 
\be \label{bc} F_{GS}=\int_0^\infty\, \frac{d\beta}{\beta}\,f(\beta)\, \exp{(-\beta E)}\equiv f \circ F_{BG},\ee 

where $f(\beta)$ satisfies 
\be \label{normita} \int_{-\infty}^{\infty}\, d\beta \, f(\beta) =1.\ee 

Note that the $\circ-$sign is used to denote the multiplicative convolution between two PDs: multiplicative convolution of two PDs $f_{X}$ and $f_{Y}$ is the PD of the product of the two corresponding random variables $X$ and $Y$.

As stated above, $\beta$ is the inverse temperature, but the integration variable may also be any convenient intensive parameter.  Superstatistics would arise as a ``statistics of statistics" that takes into account fluctuations of such intensive parameters.

\nd Beck and Cohen also show that, if $f(\beta)$ is a $\chi^2$ distribution, nonextensive thermostatistics (NEXT) is obtained, which is of interest because  
  NEXT, or Tsallis' thermostatistics, is today a very active field, perhaps a new paradigm for statistical mechanics, with applications
 to several scientific disciplines \cite{gellmann,lissia,fromgibbs}. 
 In working in a NEXt framework  one has to deal with power-law distributions, 
which are certainly ubiquitous in physics (critical phenomena
are just  a conspicuous example \cite{goldenfeld}). Indeed, it is well
known that  power-law distributions arise quite naturally
  in maximizing Tsallis'  information measure   ($q$ is a real parameter called the ``nonextensivity index") 
\be
H_{q}\left(  f\right)  =\frac{1}{1-q}\left(  1-\int_{-\infty}^{+\infty}%
f(x)^{q} dx\right), \label{dino} \ee  subject to appropriate constraints. In the case of the canonical distribution there is only one constraint, the energy $E$, i.e.,  
$\langle E\rangle = K\,\,( {\rm K\,\,a\,\,\, constant})$ and  the equilibrium
 canonical distribution writes $f(x)
=(1/Z_q)[1-(1-q)\beta_q E]^{\frac{1}{q-1}}$, with $\beta_q$ and  $Z_q$ standing for the NEXT counterparts of $\beta$ and $Z_{BG}$ above. It is a classical result that
as $q\rightarrow1,$   Tsallis entropy reduces to
Shannon entropy%
\be \label{shannon}
H_{1}\left(  f\right)  =-\int_{-\infty}^{+\infty}f(x)\log f(x).
\ee

\nd Tsallis' PDs are encountered in analyzing a rather large variety of physical systems \cite{b1,b2,b3,b4,b5}, which encourages people to continue investigating the nonextensive formalism along multiple viewpoints  and a multitude of paths. 

\vskip 4mm
\nd In such a spirit, and further pursuing along the road first travelled 
by Beck and Cohen \cite{super}, we show here that, for a fixed temperature $T$ (respectively, any adequate intensive parameter $\tau$), {\it there exists a mapping between power law PD's on the one hand, and exponential or Gaussian PD's on the other one }. Using this mapping one can transform a Tsallis PD into a Gaussian (resp., exponential) PD and vice-versa via multiplicative convolution with a chi random variable of variance unity.    

\nd This mapping nitidly reveals, for the first time as far as we know,  the (perhaps surprising) intimate relation that exists between the conventional Shannon-Gibbs-Boltzmann statistics and NEXT. As a first application we will tackle the vexing (for NEXT) problem that revolves around thermodynamics zero-th law that was first revealed by Raggio and Guerberoff \cite{01} (see also \cite{02,03,04,05,06}, a by no means exhaustive list, and references therein).

\section{Mathematical framework}

\subsection{Case $q>1$}

Let us consider a random $k-$variate vector $X=(x_1,\ldots,x_k)^{T}$ following a Tsallis
distribution with parameter $q>1$ (or $q<0$). Its distribution writes
\be
f_{X}\left(X\right) \propto \left(1-X^{T}X\right)_{+}^{\frac{n-k}{2}-1}
\label{eq:studentr}
\ee
with $n\ge k$ and with the notation $x_{+}=\max(0,x)$. As discussed in \cite{alf1}, $f_{X}$ can be obtained as the $k-$ variate marginal of a uniform distribution
on the sphere in $\mathbb{R}^{n}$. Additionally,   a stochastic representation 
for $X$ can be given in terms of Gaussians, which is essential for our present purposes, i.e., 
\be \label{siete}
X=\frac{\tilde{N}}{r}\ee
where $N=[N_{1},\dots,N_{n}]^{T}$ is a Gaussian $n-$variate vector, $\tilde{N}$ is the $k-$variate vector composed of the first $k$ components of $N$ and $r=\sqrt{\sum_{i=1}^{n}N_{i}^{2}}$
is a chi random variables with $n$ degrees of freedom. We note that
the first $k$ components of $N$ are shared by the numerator and the
denominator.
 
\begin{thm*}
\label{thm:thm1} If $a_{n}$ is i) a chi random variable with $n$ degrees
of freedom and ii) independent of a vector $X$ distributed as in (\ref{eq:studentr}),
then the product \begin{equation}
a_{n} X\sim Z\label{eq:Z}\end{equation}
where $Z$ is Gaussian with identity covariance and symbol $\sim$ denotes equality in distribution.
\end{thm*}
\begin{proof}
See the Annex.
\end{proof}
\nd Thus a Tsallis system 
can be ''Gausssianized''
simply by multiplying 
each of its components by an independent, scalar chi random
variable with $n$ degrees of freedom. For statistical mechanics purposes, 
 the variance is a very important quantity because it is proportional to the temperature \cite{pathria}.  As defined by (\ref{eq:studentr}), the random vector
$X$ has covariance matrix $R_{X}=\frac{1}{n}I_{k},$ with $I_k$ the $(k \times k)$-unit matrix.  Thus, the  random
vector $Z$, as defined by (\ref{eq:Z}) has covariance matrix (we use the notation $E X\equiv \langle X \rangle$)  \be \label{nueve}
EZZ^{T}=Ea_{n}^{2}EXX^{T}=n\times\frac{1}{n}I_{k}=I_{k}.\ee

\subsection{Case $\frac{k}{k+1}<q<1$}

Let us  first underline a useful connexion between the instances $q<1$ and $q>1$:

\begin{lem*}
if $X$ is distributed according to (\ref{eq:studentr}) then\be \label{diez}
Y=\frac{X}{\sqrt{1-X^{T}X}}\ee
is distributed as follows:\begin{equation}
f_{Y}\left(Y\right)\propto\left(1+Y^{T}Y\right)^{-\frac{m+k}{2}}\label{eq:studentt}\end{equation}
with $m=n-k$, and thus is a $k-$variate Tsallis random vector with parameter $\frac{k}{k+2}<q<1$.
A stochastic representation of $Y$ writes  \be \label{doce}
Y=\frac{N}{\chi_{m}},\ee
where $N$ is a $k-$variate Gaussian vector and $\chi_{m}$ is
a chi random variable independent on $N$ with $m=n-k$ degrees
of freedom. Note moreover that the covariance matrix of $Y$ writes \be \label{trece}
EYY^{T}=ENN^{T}E\frac{1}{\chi_{m}^{2}}=\frac{1}{m-2}I_{k}.\ee

\end{lem*}
\nd Now we can state our main theorem:

\begin{thm*}
\label{thm:thm2}if $Y$ is distributed according to (\ref{eq:studentt})
and if $a_{n}$ is a chi distributed random variable with $n$ degrees
of freedom independent on $Y$, then random variable  \be \label{catorce}
Z=\frac{a_{n}}{\sqrt{1+Y^{T}Y}} Y\ee
is Gaussian with unit covariance. Moreover, random variable \be \label{quince}
b_{m}=\frac{a_{n}}{\sqrt{1+Y^{T}Y}},\ee
is chi distributed with $m$ degrees of freedom. Note that $b_{m}$
is dependent on $Y.$
\end{thm*}
\begin{proof}
\nd See the Annex.
\end{proof}

\subsection{Extension of the above theorems}

The precedent considerations can be extended from the Gaussian case to the one of a  purely exponential factor. A
 given Tsallis-distributed vector $X$ gets transformed, via multiplicative convolution with a chi random 
 variable, into a vector that is distributed according to an exponential PD.  In other words, 
 from a Gaussian vector $G$, one can  
 create a Tsallis distributed vector $X$ by multiplying by a random scalar variable. This mapping is invertible:
  a Gaussian distributed vector can be obtained by multiplying a Tsallis distributed one by another properly chosen random variable.

\subsection{Application: the merging of two independent systems}

\nd Suppose we have two independent systems (in the sense that their mutual interaction is negligible) whose states are described
by two  Tsallis random vectors, independent as well,  $X$ and $Y$ with, say,  $q>1$.
If we consider the system $Z^{T}=\left[X^{T},Y^{T}\right]$ we immediately realize that it 
 is not Tsallis distributed since its distribution writes \be \label{16} 
f_{Z}\left(X,Y\right)=\left(1-X^{T}X\right)^{\frac{n-k}{2}-1}\left(1-Y^{T}Y\right)^{\frac{n-k}{2}-1},\ee
and, consequently,  can not be expressed as a function of $X^{T}X+Y^{T}Y$ (except
in the Gaussian case $q=1$) \cite{01,02,03,04,05}.

\vskip 3mm
\nd The problem can, however, be circumvented as follows: if we now ''pre-multiply'' system $X$ (resp. $Y$) with the same
chi-distributed random variable $a_{n}$ as defined in theorem \ref{thm:thm1},
then the system 

\be \label{17} Z=a_{n} \left[X^{T},Y^{T}\right]^{T} \ee is the merging
of two Gaussian systems. Moreover, the covariance of $Z$ writes \be  \label{18}
EZZ^{T}=I_{2k},\ee
so that $a_{n} X$ and $a_{n} Y$ are both uncorrelated and Gaussian, and thus
independent. Finally, considering a new chi distributed random variable $b_{m}$ with $m$ degrees of freedom and independent of $Z$, the Tsallis distributed vector 
\be \label{19} X_{Tsallis} = \frac{Z}{b_{m}}=\frac{a_{n}}{b_{m}}  \left[X^{T},Y^{T}\right]^{T}, \ee 
- and NOT (\ref{16}) - turns out to be the ``true" representative of the ``merging" of two nonextensive systems. Such a vector is not characterized by a fixed temperature $T$ (or, more generally, by a fixed value of an appropriate intensive system's parameter $\tau$), but by a superposition of temperatures ``centered" at $T$ (resp., superposition of the intensive parameter centered at $\tau$) , exactly in the spirit of superstatistics. We can then speak of a ``stochastic" zero-th law that would apply for nonextensive thermostatistics: {\it given two independent systems $A,\,\,B$ of equal temperature $T$, the pertinent temperature of the associated composite system $A+B$ at equilibrium fluctuates around $T$.}

\section{Physical considerations}

\nd For didactic reasons it is convenient at this point to briefly review the well-known connection between variance and temperature for a Gaussian PD before probing further into the zero-th law.

\subsection{Quadratic classical Hamiltonians}

\nd We consider first the ``normal modes" hamiltonian for $N$ independent particles in three dimensions

\be \label{qch} H=\frac{1}{2}\sum_{m=1}^{3N} c^q_m q_m^2 +  c^p_m p_m^2\equiv \sum_{i=1}^{6N} c_i z_i^2/2, \ee
and consider the location of the many-body system in its $6N-$dimensional $\Gamma-$space. This is represented by a $6N-$vector. Obviously, $\Gamma-$position has probabilities proportional to a function of $\beta H$. We consider two instances: conventional BG thermostatistics and Tsallis' one.
\begin{itemize} \item For  BG statistics,  location is represented by a vector $Z$ distributed according to the $(1/Z)\exp{(-\beta H)}$ PD. On account of (\ref{qch}) then, $Z$ is Gaussian-distributed.
 \item For Tsallis' statistics, locations is represented by a vector $X$ that is Tsallis-distributed.
\end{itemize}
If $Z=(z_1,\ldots,z_{6N})^{T}$ is BG-distributed, then 
its covariance matrix $EZZ^T$ is a diagonal matrix. Because of the equipartition theorem \cite{reif}, the expectation value of each member of the diagonal equals $k_BT/2$. Thus, $EZZ^T=(k_BT/2)I_{6N}$, and the covariance matrix is proportional to the temperature.

\nd In the case of Tsallis statistics, let the vectors in (\ref{16}) be distributed according to 

\ben \label{20} f_X(x) &\propto& [1- (1-q) \beta \sum_{i=1}^k x_i^2] \cr  f_Y(y) & \propto & [1- (1-q) \beta \sum_{i=1}^k y_i^2],\een so that we are merging two system of inverse temperature $\beta= 1/k_BT$. The representative of the merged system is (\ref{19}). Note that $Z$ there is a Gaussian vector of temperature $T$, and $X_{Tsallis}$ a Tsallis vector that, although not having a definite temperature, possesses a temperature distribution ``peaked" at $T$. The zero-th law of thermodynamics for Tsallis distributed vectors entails that if we join two nonextensive systems of temperature $T$ and wait until equilibrium is re-established, we obtain a total system whose temperature ``fluctuates" (\`a la Beck-Cohen) around this temperature $T$. 

\subsection{``Normal modes" or diagonalizable Hamiltonians}

Our approach can be extended to more general types of Hamiltonians of the type

\be H= \sum_{i=1}^{6N} c_i z_i^{p_{i}},\,\,\,p_{i}\,\,an\,\, integer\,\, power,  \ee 
provided the mixing variable $a_{n}$ is modified as follows.

\begin{thm*}
\label{thm:thm3} If $A$ is a diagonal matrix whose element $A_{i,i}=a^{\frac{1}{p_{i}}}$ where $a$ is i) a chi random variable with $n-k+2\sum_{i=1}^{k}\frac{1}{p_{i}}$ degrees of freedom and ii) independent of component  $X_{i}$ distributed as in (\ref{eq:studentr}),
then the product 
\begin{equation}
A X\sim Z
\label{eq:Z_A}
\end{equation}
where $Z$ is distributed as 
\be
f_{Z}(Z)\propto \exp \left(  -\sum_{i=1}^{k} c_{i}z_{i}^{p_{i}}\right).
\ee
\end{thm*}
\begin{proof}
\nd See the Annex.
\end{proof}

\nd Following now an identical line of reasoning as that of  Subsection 2.4, we can generalize the stochastic zero-th law obtained there to all classical systems described by 
the above type of  Hamiltonian.

\section{Conclusions}

\nd We have here generalized the concept of superstatistics devised by Beck and Cohen \cite{super}. Its typical ``convolution path" that involves fluctuations of intensive parameters allows one to express generalized  factors $B_{GF}$ in terms of a superposition of exponential PDs. We show the the path can be   ``inverted", at least with reference to Tsallis PDs, in the sense that a normal (or an exponential) distribution can also  be cast as a superposition of Tsallis PD. {\it Stochastic invertible mappings between power law distributions and Gaussian (or exponential) ones have been introduced and discussed}.

\nd As an application we have shown that the zero-th  law of thermodynamics is obeyed by Tsallis PDs only in a stochastic sense. In nonextensive thermostatistics, if two  systems of temperature $T$ are merged, the ensuing equilibrium total distribution exhibits fluctuation around that temperature.

\section{Annex: Proofs}

\subsection{Proof of theorem \ref{thm:thm1}}

>From \cite[3.471.3]{Gradshteyn}, we have
\be
\exp\left(-cx\right)=\frac{c^{\alpha}}{\Gamma\left(\alpha\right)}\int_{0}^{+\infty}\exp\left(-c/u\right)u^{-\alpha-1}\left(1-ux\right)_{+}^{\alpha-1}du 
\label{eq:general}
\ee

Now replace $x$ by $\sum_{i=1}^{k}x_{i}^{2}$ and $u$ by $v^{-1}$
in (\ref{eq:general}) to obtain
\begin{eqnarray*}
\exp\left(-c\sum_{i=1}^{k}x_{i}^{2}\right) & = & \frac{c^{\alpha}}{\Gamma\left(\alpha\right)}\int_{0}^{+\infty}\exp\left(-cv\right)v^{\alpha-1}\left(1-v^{-1}\sum_{i=1}^{k}x_{i}^{2}\right)_{+}^{\alpha-1}dv\\
 & = & \frac{c^{\alpha}}{\Gamma\left(\alpha\right)}\int_{0}^{+\infty}\exp\left(-cv\right)v^{\alpha-1+\frac{k}{2}}\left(\frac{1}{\sqrt{v}}\right)^{k}\left(1-\sum_{i=1}^{k}\frac{x_{i}^{2}}{v}\right)_{+}^{\alpha-1}dv.
\end{eqnarray*}
Now, as a classical Gamma integral, 
\[
\int_{0}^{+\infty}\exp\left(-cv\right)v^{\alpha-1+\frac{k}{2}}dv=\frac{\Gamma\left(\alpha+k/2\right)}{c^{\alpha+k/2}}
\]

so that  

\[
f_{v}\left(v\right)=\frac{c^{\alpha+k/2}}{\Gamma\left(\alpha+k/2\right)}\exp\left(-cv\right)v^{\alpha-1+\frac{k}{2}}\]
is the distribution of a chi-squared random variable with $k+2\alpha$
degrees of freedom and variance $\sigma^{2}=1/2c$. Thus
\begin{eqnarray*}
\left(\frac{1}{2\pi\sigma^{2}}\right)^{k/2}\exp\left(-\frac{\sum_{i=1}^{k}x_{i}^{2}}{2\sigma^{2}}\right) & = & \frac{c^{\alpha}\Gamma\left(\alpha+k/2\right)}{c^{\alpha+k/2}\left(\sigma\sqrt{2\pi}\right)^{k}\Gamma\left(\alpha\right)}\int_{0}^{+\infty}\sigma\chi_{2\alpha+k}\left(v\right)\left(\frac{1}{\sqrt{v}}\right)^{k}\left(1-\sum_{i=1}^{k}\frac{x_{i}^{2}}{v}\right)_{+}^{\alpha-1}dv.\\
 & = & \int_{0}^{+\infty}\sigma\chi_{2\alpha+k}\left(v\right)S_{X\sqrt{v}}^{\left(\alpha\right)}\left(x_{1},\dots,x_{k}\right)dv\end{eqnarray*}
where $S_{X}^{\left(\alpha\right)}$ denotes a $k-$variate Tsallis
distribution obtained as the marginal of the uniform distribution
on the sphere $S_{n}$ \[
S_{X}^{\left(\alpha\right)}\left(x\right)=\frac{\Gamma\left(\alpha+k/2\right)}{\left(2\pi\sigma^{2}c\right)^{\frac{k}{2}}\Gamma\left(\alpha\right)}\left(1-x^{T}x\right)^{\alpha-1}\]
where\[
\frac{n-k}{2}=\alpha\leftrightarrow n=2\alpha+k.\]

\subsection{Proof of theorem \ref{thm:thm2}}

We know from the lemma that \[
Y=\frac{X}{\sqrt{1-X^{T}X}}\]
and we deduce that \[
X=\frac{Y}{\sqrt{1+Y^{T}Y}}.\]
>From theorem \ref{thm:thm1}, we know that $a_{n}X$ is normal. Thus\[
\frac{a_{n}}{\sqrt{1+Y^{T}Y}}Y\]
is normal. Moreover, the distribution of random variable $b_{m}=a_{n}/\sqrt{1+Y^{T}Y}$
can be computed as follows\begin{eqnarray*}
f_{b}\left(b\right) & \propto & \int\left(1+Y^{T}Y\right)^{n/2}b^{n-1}\exp\left(-\frac{b^{2}\left(1+Y^{T}Y\right)}{2}\right)\left(1+Y^{T}Y\right)^{-\frac{m+k}{2}}dY\\
 & = & b^{n-1}\exp\left(-\frac{b^{2}}{2}\right)\int\exp\left(-\frac{b^{2}Y^{T}Y}{2}\right)dY\\
 & \propto & b^{n-1}\exp\left(-\frac{b^{2}}{2}\right)b^{-k}=b^{m-1}\exp\left(-\frac{b^{2}}{2}\right)
\end{eqnarray*}
so that $b_{m}$ is a chi random variable with $m$ degrees of freedom.

\subsection{Proof of theorem \ref{thm:thm3}}
Starting form equality (\ref{eq:general}), we obtain

\begin{eqnarray*}
\exp\left(-c\sum_{i=1}^{k}x_{i}^{p_{i}}\right) &=&
\frac{c^{\alpha}}{\Gamma\left(\alpha\right)}\int_{0}^{+\infty}\exp\left(-cv\right)v^{\alpha-1}
\left(1-\sum_{i=1}^{k} 
\left(\frac{x_{i}}{v^{\frac{1}{p_{i}}}}\right)^{p_{i}}\right)_{+}^{\alpha-1}dv \\
& = & 
\frac{c^{\alpha}}{\Gamma\left(\alpha\right)}\int_{0}^{+\infty}\exp\left(-cv\right)v^{\alpha-1+\frac{1}{p}}
\left(\prod_{i=1}^{k}\frac{1}{v^{\frac{1}{p_{i}}}}\right)
\left(1-\sum_{i=1}^{k} 
\left(\frac{x_{i}}{v^{\frac{1}{p_{i}}}}\right)^{p_{i}}\right)_{+}^{\alpha-1}dv
\end{eqnarray*}

with $\frac{1}{p}=\sum_{i=1}^{k} \frac{1}{p_{i}}$. Following the same path as in proof of theorem \ref{thm:thm3}, we deduce that vector with $i-$th component
$a^{p_{i}}X_{i}$, where $X_{i}$ is distributed as
\be
f_{X_{i}}(x)=\left(1-x^{p_{i}}\right)_{+}^{\alpha-1},
\ee
is distributed as vector $Z$ with

\be
f_{Z}(Z)\propto \exp \left(  -\sum_{i=1}^{k} c_{i}z_{i}^{p_{i}}\right)
\ee

\end{document}